\documentclass[conference]{IEEEtran}
\IEEEoverridecommandlockouts
\usepackage{cite}
\usepackage{amsmath,amssymb,amsfonts}
\usepackage{graphicx}
\usepackage{textcomp}
\usepackage{xcolor}

\usepackage{color,xspace,listings,comment,paralist,tabulary,svg}
\usepackage{algorithm,algpseudocode}%
\usepackage[colorinlistoftodos]{todonotes}

\usepackage{url}
\usepackage{booktabs}

\def\BibTeX{{\rm B\kern-.05em{\sc i\kern-.025em b}\kern-.08em
    T\kern-.1667em\lower.7ex\hbox{E}\kern-.125emX}}

\begin{document}

\title{\hirace: Accurate and Fast \\ Source-Level Race Checking of GPU Programs
}

\author{\IEEEauthorblockN{John Jacobson}
\IEEEauthorblockA{\textit{Kahlert School of Computing} \\
\textit{University of Utah}\\
Salt Lake City, USA \\
john.jacobson@utah.edu}
\and
\IEEEauthorblockN{Martin Burtscher}
\IEEEauthorblockA{\textit{Department of Computer Science} \\
\textit{Texas State University}\\
San Marcos, USA \\
burtscher@txstate.edu}
\and
\IEEEauthorblockN{Ganesh Gopalakrishnan}
\IEEEauthorblockA{\textit{Kahlert School of Computing} \\
\textit{University of Utah}\\
Salt Lake City, USA \\
ganesh@cs.utah.edu}
}

\newcommand\ggcmt[1]{\todo[inline, size=\small, color=green!40]{GG: #1}}
\newcommand\ggcmtside[1]{\todo[size=\scriptsize, color=green!40]{GG: #1}}

\newcommand\jjcmt[1]{\todo[inline, size=\small, color=blue!60]{JJ: #1}}
\newcommand\jjcmtside[1]{\todo[size=\scriptsize, color=blue!40]{JJ: #1}}

\newcommand\mbcmt[1]{\todo[inline, size=\small, color=red!60]{MB: #1}}
\newcommand\mbcmtside[1]{\todo[size=\scriptsize, color=red!40]{MB: #1}}

\newcommand{\hirace}{{\textsc HiRace}\xspace}
\newcommand{\gklee}{{\textsc GKLEE}\xspace}
\newcommand{\gpuverify}{{\textsc GPUVerify}\xspace}
\newcommand{\pug}{{\textsc PUG}\xspace}
\newcommand{\curd}{{\textsc CURD}\xspace}
\newcommand{\barracuda}{{\textsc Barracuda}\xspace}
\newcommand{\iguard}{{\textsc iGUARD}\xspace}
\newcommand{\ld}{{\textsc LD}\xspace}
\newcommand{\cudamemcheck}{{\textsc Compute Sanitizer}\xspace}
\newcommand{\fasttrack}{{\textsc FastTrack}\xspace}
\newcommand{\scord}{{\textsc ScoRD}\xspace}
\newcommand{\sword}{{\textsc SWORD}\xspace}
\newcommand{\archer}{{\textsc Archer}}
\newcommand{\tsan}{{\textsc ThreadSanitizer}\xspace}

\newcommand{\syncthreads}{\texttt{\_\_syncthreads}}
\newcommand{\syncwarp}{\texttt{\_\_syncwarp}}

\newcommand{\fsm}{FSM}

\maketitle

\begin{abstract}

Data races are egregious parallel-programming bugs on CPUs. They are even worse on GPUs due to the hierarchical thread and memory structure, which makes it possible to write code that is correctly synchronized within a thread group while not being correct across groups. 
%
%GPU programs that contain data races are unreliable for most tasks.\ggcmtside{simple good way i think - not shrill}
%
Thus far, all major data-race checkers for GPUs suffer from {\em at least one}
of the following problems:
%()~
%cannot handle data-dependent control flow,
%(1)~
they do not check races in global memory,
%(2)~
do not work on recent GPUs,
%(3)~
scale poorly,
%(4)~
have not been extensively tested,
%(5)~
miss simple data races,
%(6)~
or are not dependable without detailed knowledge of the compiler.

Our new data-race detection tool, \hirace, overcomes these limitations.
Its key novelty is an innovative parallel finite-state machine that condenses an arbitrarily long access history into constant-length state, 
thus allowing it to handle large and long-running programs.
\hirace is a dynamic tool
%(handles control-flow naturally),
that checks for thread-group shared memory and global device memory
races. It utilizes source-code instrumentation, thus avoiding 
driver, compiler, and hardware dependencies.
We evaluate it on a modern calibrated data-race benchmark suite. On the 580 tested CUDA kernels, 346 of which contain data races, \hirace finds races missed by other tools without false alarms and is more than 10 times faster on average than the current state of the art, while incurring only half the memory overhead.
%
%To produce a more meaningful evaluation, we rely on a unique approach where we employ a benchmark-suite generator of graph algorithms that can generate very large irregular graph codes that have control-flows and systematically mutates any of these codes to introduce a  variety of known races.
%
% and 6 graphs ranging between 5 nodes and 200 nodes (numbers which the best alternate race-checker can handle), we show that \hirace OUTPERFORMS.
%
\end{abstract}


\section{Introduction}
\label{sec:intro}
%\input{sections/misc}

%\ggcmt{vector clocks \#$thread^{2} + c \cdot$ memsize - a point to make somewhere; rest is handled in my writing.}
% \begin{itemize}
%     \item GPU importance (deep learning, hpc, applications in general..)
%     \item Why is race detection different on gpu?
%     \item massive parallelism, 
%     \item weaknesses of existing methods (vector clock probs)
%     \item vector clocks \#$thread^{2} + c \cdot memsize$
% \end{itemize}

Data races are notoriously difficult to identify by manual code inspection or through the use of unit tests due to the complexity of modern memory models and the non-deterministic nature of race conditions.
Data races are known to change results when ported to new platforms~\cite{gklee} and fundamentally produce unreliable binaries~\cite{jack-about-boehm-adve} because the assumptions under which a compiler optimizes code are invalidated~\cite{boehm2011benign} by data races. 
GPU programs are in many ways more susceptible to data races due to the large number of concurrent threads and the availability of multiple layers of shared memory through which these threads may interact. Although GPU languages provide synchronization primitives and atomics to enforce orderings on memory actions, their use is not always intuitive and they have significant performance costs. Developers often utilize complex code patterns to avoid synchronization overheads, making the code difficult to reason about. This has resulted in many attempts at creating GPU data-race detection tools (see Table~\ref{tab:gpu-race-checkers}), yet there are still no good practical solutions available even for the most widely used GPU languages. 

Most of the GPU data race checkers developed to
date in Table~\ref{tab:gpu-race-checkers} are not in working condition,
thus leaving the community with two options (both of which are specific to NVIDIA GPUs), namely \iguard~\cite{iguard} and \cudamemcheck~\cite{cuda-memcheck-url}.
However, \cudamemcheck~does not detect global memory races---a serious omission considering that many GPU programs concurrently access global memory to reduce data movement.
\iguard, and its future evolution,
heavily depends on support by the NVBit tool~\cite{nvbit}, does not seem to be actively maintained, and is not functional on NVIDIA architectures released since its publication. 

\begin{table*}[t]
\begin{center}
\begin{tabulary}{1.0\textwidth}{LLL}
\toprule
\textbf{Tool Name} & \textbf{Status} & \textbf{Generality, Usability} \\
\midrule
\gpuverify~\cite{betts2012gpuverify}
& Not Maintained & Static analyzer, limited applicability\\
%\hline 
\gklee~\cite{gklee} & Not Maintained & CUDA as of 2016 \\
%\hline 
\pug~\cite{pug} & Not Maintained & Limited CUDA subset \\
%--
\curd~\cite{CURD} & Not Released & CUDA \\
%\hline 
\barracuda~\cite{eizenberg2017barracuda} & Not Maintained & CUDA \\
%\hline 
\ld~\cite{ld-ding} &
Never Released &
Useful when tracking states affordable \\
%\hline 
\iguard~\cite{iguard} & Not Maintained & NVIDIA SASS binary instrumentation\\
%
%\hline 
%--
Faial~\cite{faial-drf:fmsd23} & Recent release &
Limited to a dozen or so straight-line 
code blocks
without conditionals or loops \\
\cudamemcheck~\cite{cuda-memcheck-url} & Maintained & NVIDIA proprietary,  No global-memory data race detection \\
\bottomrule
\end{tabulary}
\end{center}
\caption{Existing GPU Data-Race Checking Tools}
\label{tab:gpu-race-checkers}
\end{table*}

In this paper, we present \hirace{}, a practical and efficient dynamic data-race checker that aims to serve the needs of a vast majority of today's HPC and ML practitioners. \hirace{} works at the source-level by automatically instrumenting code through Clang's source rewriting API~\cite{lattner_llvm_2011}, part of the widely adopted and actively maintained LLVM project. Source-level instrumentation allows us to avoid reliance on compiler contracts and variability due to compiler optimizations.
Functionally, \hirace{} is designed with a focus on the most common GPU programming patterns, providing significant improvements in completeness, run time, and memory overhead for codes utilizing GPU synchronization primitives. 

\vspace{1ex}
\noindent{\bf Design Novelties} 
\vspace{1ex}

{\sl Single Access Record:\/} The most significant novelty in \hirace is based upon the observation that a fixed-size accessor record
(elaborated in Section~\ref{sec:design})
is sufficient to encode all necessary information about {\em all} prior accesses to identify races.
%, which we encode in our new state machine.
%
This design ensures that the memory overhead is constant and small.
In comparison, other designs maintain multiple accessor records to dynamically reconstruct the access history. This requires significantly more memory, and repeated accesses to a single location can cause an accessor record to be evicted due to capacity conflicts, causing race omissions (discussed further in Section~\ref{sec:eval}).

\vspace{1ex}
{\sl Explicit State-Machine:\/} \hirace{} achieves a modest size (constant space per memory location) access history through the use of an explicit finite-state machine that encodes access histories compactly (5 bits of state per tracked memory location). This is achieved in part by equating threads to thread groups following the GPU concurrency hierarchy as elaborated in Section~\ref{sec:design}.

\vspace{1ex}
{\sl Correctness:\/} Having an explicit state-machine also allowed us to develop a Murphi~\cite{Dill1996TheMV} formal reference model, which we tested with a model-checker to match against \hirace's behavior.
This exercise gives increased confidence in the correctness of our design, and aids in extending our methodology to new and more complex memory models (a similar verification exercise was reported in~\cite{verified-race-checking-wilcox-flanagan}).

To further increase our confidence in
\hirace's race-checking completeness,  we test it against {\em calibrated} benchmarks from the Indigo suite~\cite{mb-ispass22}, which contains thousands of GPU codes that are labeled as bug-free or buggy, including which bugs they contain.
Indigo benchmark codes contain irregular graph algorithms, which exhibit data-dependent memory accesses and control flow.
Such behavior makes programs harder to debug because even buggy codes
will execute correctly for inputs that happen to yield (1) control flow that avoids the problematic
code sections or (2) memory-access patterns that exclude the problematic data dependencies. In
other words, only certain inputs may trigger the software defects present in the code. Moreover,
the thread timing in parallel programs similarly only triggers software defects in some but not all
executions of a program, even when using the same input. Together, this makes detecting bugs in
irregular parallel programs particularly challenging, and we believe makes a suite like Indigo an ideal test target for verification tools like \hirace.

We observe that \hirace's reporting conforms to Indigo's calibration and also identified an unintended data race within some Indigo benchmarks.
%
% Another valuable aspect of our study is
% that we can use Indigo to generate codes which,
% for the same algorithm, employ different concurrency patterns.
% %
% This allows us to push on scale.
%
We find that {\bf \hirace can be 30$\times$ to 50$\times$ faster than state of the art---with median speedups of at least 7$\times$} (Figure~\ref{fig:speedup_iguard}).

\vspace{1ex}
 This paper makes the following main contributions. \hirace
 \begin{compactitem}

 \item employs an innovative state-machine-based design,

 \item detects more races than
 the state of the art in dynamic GPU data-race detection,

 \item is on average 10$\times$ faster and scales to much larger programs compared to state of the art,

 \item has been tested against a calibrated set of benchmarks,
 
 \item uses source-level instrumentation based on Clang, and
 
 \item has been verified with the Murphi model checker.
\end{compactitem}

\hirace{} will be submitted to the artifact committee and will be open-sourced after the end of the anonymity phase.

\vspace{1ex}
Section~\ref{sec:background} provides
background on shared-memory
race checking in general and
 GPU data-race checking in particular.
 Section~\ref{sec:design} presents a systematic walk-through of \hirace's design.
 Section~\ref{sec:impl} describes our experimental
 methodology.
 Section~\ref{sec:eval} presents our result evaluation.
 Section~\ref{sec:concl-remarks} provides concluding remarks.

\section{Background}
\label{sec:background}
Data-race checking is a topic with a long history~\cite{netzer1989detecting}. 
In this section, we outline important background on data races and CUDA programming paradigms. We also present prior work and contrast it with \hirace.

\subsection{Data Races}

A  parallel program has a data race if multiple threads access the same memory location, at least one of the accesses is a write, and the accesses are concurrent (more precisely, are not ordered by some {\em happens-before}~\cite{Lamport79} relation).
The definition of a data race has been refined to its present form based on a precise notion
of happens-before~\cite{jack-about-boehm-adve}, which is supported by venerable tools such as \fasttrack~\cite{flanagan_fasttrack:_2009} and Google's Thread Sanitizer~\cite{serebryany_threadsanitizer:_2009}---primarily addressing shared memory (POSIX-like) threads.

Consequently, detecting data races requires first identifying all memory accesses, particularly those that are {\em conflicting} (accesses to the same location from different threads), and second determining whether any conflicting accesses thus identified are concurrent.
While doing so, data-race-detection tools aim for a balance between two properties: reporting all races (``no omissions'', or {\em completeness}) and NOT reporting any non-races  (``no false alarms'', or {\em soundness}).
There are two primary approaches to making these determinations: \emph{static} and \emph{dynamic} analysis of a program.

Static analysis is performed by parsing program instructions without executing them. This generally involves utilizing some encoding of the program semantics that model the relevant behaviors of the code being analyzed. 
Static analyses have the benefit of complete program knowledge, but, unfortunately, suffer from the undecidability of many program properties (for example, the halting problem). As a result, static analysis methods inherently suffer from reporting false positives. This is a significant issue in practice as it may add cost either in the form of code review time or unnecessary development time and overhead in programs, and they may erode confidence in the tool's other error reports, causing users to ignore real errors.
For this reason, our work on \hirace focuses on dynamic analysis.

Dynamic analysis is performed by monitoring the underlying program at run-time. Doing so avoids the burden of determining reachability of instructions since all executed instructions can be directly witnessed.
However, dynamic methods are restricted to analyzing program instructions that are observed in a given test execution. Any conditional code paths that are not explored by a given configuration cannot be tested, leaving regions of code unanalyzed. This problem can be alleviated by testing the code with multiple inputs that elicit the various possible program behaviors.
%
%We intend to address this weakness of dynamic methods in future work by adding separate static analysis components to increase automatic code coverage.

Many modern tools employ a combination of these methods, such as Archer~\cite{atzeni_archer:_2016} and GKLEE~\cite{DBLP:conf/ppopp/LiLSGGR12} (discussed below).

\subsection{CUDA Shared-Memory Concurrency}

CUDA devices present large numbers of threads to the user, which are organized into hierarchical structures to be dynamically scheduled and executed on a number of streaming multiprocessors (SMs). Each SM has several SIMT vector units. Modern GPUs have over 100 SMs on a single device, each capable of hosting over 1,000 resident threads. 

Logically, individual threads are arranged into \emph{thread blocks}, which are executed on a single SM. The threads of a thread block are arranged into \emph{warps}, which are executed on a single SIMT unit. 
Device memory is composed of several spaces (or \emph{scopes}) with differing accessibility. \emph{Global memory} is accessible by all threads on the device, whereas \emph{shared memory} is allocated to individual thread blocks and is accessible only to threads within the same thread block. 

For managing this thread hierarchy, CUDA provides a set of block-scope synchronization primitives (\emph{barriers}) as well as a set of warp-scope synchronization primitives.

The most basic block scoped barrier is \syncthreads. Informally, any thread that executes a \syncthreads{} instruction will wait for all threads within the same thread block to execute the same \syncthreads{} before proceeding to its next instruction.
There are variants of this primitive that provide the same guarantee but also perform a thread-block-wide reduction operation and broadcast the result to all threads in the block.

The most basic warp-scoped barrier is \syncwarp. It provides the same functionality as \syncthreads{} but is limited to the threads within the same warp. Additionally, \syncwarp{} takes a \emph{mask} argument that restricts the participating threads to only those specified in the mask (which must all be members of the same warp).
Like for the block-level barriers, there are additional warp synchronization primitives that allow thread communication alongside the synchronization effects.

A barrier-only concurrency structure without data-dependent conditionals enjoys the so called 
{\em two-thread theorem}~\cite{pug,DONALDSON20173}, which is employed
in \gklee~and \gpuverify. It states that, if there is a race, it can be discovered within the scope of just two threads.
%
% When analyzing programs using only these natural concurrency patterns, GPU race checkers have the property that, if the code contains a data race, this race can be identified under any schedule (i.e., they are {\em predictive}~\cite{bond-predictive} by nature).
% \mbcmt{please rephrase the above paragraph, it is hard to understand/read}

\subsection{Other Related Work}

The CPU-based \fasttrack algorithm for Java pioneered many ideas in dynamic race checking that can still be found in current checkers.
Specifically, the idea of maintaining multiple access records to each word in memory originated in \fasttrack and is utilized in \tsan{} and its derivatives, as well as in \iguard.
\fasttrack showed formally that, while writes in a data-race-free program are totally ordered, reads are not. As a result, the algorithm requires an unbounded read history for each word to be complete. In practice, the access history generally contains only 2 or 4 prior accessors in tools that implement this algorithm on the CPU. \iguard is only able to store a single prior accessor due to the size of each access record on the GPU.
In contrast, \hirace summarizes this history by tracking {\em groups of threads} as a single identifier, along with a small state tag (discussed in Section~\ref{sec:design}). This allows us to achieve a much smaller memory footprint as well as a more complete analysis since \hirace does not need to evict prior access history.

In the GPU arena, race-checkers based on static analysis (\gpuverify, \pug), dynamic analysis at the source level (\curd, \ld), dynamic symbolic execution (\gklee), and dynamic binary-level analysis (\iguard, \barracuda) have been developed, as summarized in Table~\ref{tab:gpu-race-checkers}.

\cudamemcheck{} is NVIDIA's proprietary correctness tool suite for CUDA programs and includes a \emph{RaceCheck} tool. The details of these tools' implementation are not publicly available, but since it is included within the standard CUDA toolkit, it is the most widely available and easy to use race-detection tool currently available for GPUs. Unfortunately, \cudamemcheck{} does not check global device memory for data races. Moreover, we found that it also misses data races {\em in shared memory}, which \hirace is able to identify (discussed in Section~\ref{sec:eval}).

\iguard~\cite{iguard}, an evolution from \scord, 
the prior reported in~\cite{scord}.
It is an open-source dynamic binary-instrumentation-based data-race-detection tool for CUDA programs.
To our knowledge, \iguard represents the current state of the art in GPU data-race detection as it outperforms other available tools in most categories.
It utilizes the NVBit~\cite{nvbit} API to dynamically instrument CUDA programs.
It is simple to use as the compiled binary can be loaded as a shared library at runtime when executing the test program, and it supports both global and shared memory.
The design of \iguard is based on the same logic presented in \fasttrack but optimized for execution on CUDA devices.
Unfortunately, this design relies on an unbounded access history for completeness, and, due to the scale of CUDA programs, only one prior accessor can be tracked in 16 bytes of metadata per monitored word of memory.
More importantly, the code base has not been updated since its publication and newer generations of CUDA devices are no longer supported by \iguard.
The primary distinction between \hirace{} and \iguard{} is that the latter was designed with a focus on traditional CPU synchronization patterns (particularly on automatic lock inference), which are less common on GPUs, whereas \hirace{} focuses on the most common patterns, which utilize barrier primitives or atomics.

\section{Design of \hirace}
\label{sec:design}
Our design is based around the built-in synchronization primitives available in GPUs (such as CUDA's \emph{\_\_syncthreads}) as opposed to focusing on the full generality of CPU parallel programming patterns. Our method does not preclude the handling of fine-grained thread synchronization, but we focus on optimizing the most common and impactful forms of thread communication in GPUs.

We achieve this by relating all possible thread access histories to a single memory location.
For example, if a thread reads the same memory location multiple times, we need not record each access separately. A single thread ID and a flag indicating a read access by that ID suffices.
Going a step further, in a program with only 2 threads, if both threads concurrently read the same memory location (in any order and any number of times), it suffices to only track the ID of either accessor along with a flag indicating that {\em multiple} threads have read this location.

%% always input figures - easier to move inclusion spot then
%\input{sections/fsm-blocksync}

In the general case, this would require $2^T$ flags to track all possible thread relations between $T$ accessing threads. However, due to the hierarchical and symmetric nature of the shared memory scopes of GPUs, instead of tracking individual threads at every memory interaction, {\em we only track 
the hierarchical scope shared by the accessing threads}.
These scopes provide naturally symmetric work groups that are uniquely determined by the ID of any thread within the group (i.e., certain bits of a global thread ID). As such, a single identifier (the last accessing thread ID) and a flag representing the scope of access (i.e., thread-block scope) are sufficient to characterize a large class of unbounded access histories (arbitrary reads from multiple threads within the same thread-block).

\hirace is based on a
finite state machine (\fsm) that encodes the
happens-before logic introduced earlier.
%
% (although the machine can easily be extended to other models and deeper hierarchies).
%
% The access history\ggcmtside{smoothen...} and happens-before logic within \hirace are entirely contained within a
Whereas the \fsm{}
is tailored to the current CUDA programming model,
it can easily be extended to other models and deeper hierarchies.
The \hirace{} \fsm{} has 25 states and 1200 transitions between states. They are sufficient to describe the interactions of all atomic and non-atomic reads and writes to a single memory location as well as all needed synchronization history from warp-scoped and block-scoped barrier primitives.

Rather than exhaustively detailing each state of the \fsm{} (whose complete specification will be made publicly available), this section incrementally develops similar state machines for simpler concurrency models.
To this end, we present two
scenarios that introduce the
key ideas behind our approach:
how the \fsm{} performs
atomic updates, how our
source instrumentation
works, and how we succinctly express
the \fsm{} transitions.

\subsection{Simple Scenarios}

% \vspace{1ex}
% \noindent{\bf Simple Model} 
% \vspace{1ex}
% \newline

\begin{minipage}{\linewidth}\begin{lstlisting}[caption={Read-write data race},label={lst:rw-nosync}, basicstyle=\small]
__global__ void race_noSync(int* data) {
  int i = threadIdx.x + blockIdx.x * TPB;
  int val = data[0];
  data[0] = i + val;  // race!
}
\end{lstlisting}
\end{minipage}

% \begin{minipage}{\linewidth}\begin{lstlisting}[caption={read-write data race},label={lst:rw-nosync}, basicstyle=\small]
% __global__
% void race_noSync(int* data) {
%     int i = threadIdx.x + blockIdx.x * ThreadsPerBlock;
%     int val = data[0];
%     if (i > 0) {
%         data[i-1] = i; // race!
%     }
% }
% \end{lstlisting}
% \end{minipage}

First, we consider the set of CUDA programs containing only non-atomic reads and writes to a single memory space (global memory) that use no form of synchronization. Listing~\ref{lst:rw-nosync} provides an example of such a program.

\begin{figure}[htb]
    %\centering
    %\includesvg[width=\textwidth,inkscapelatex=false,extractformat={pdf,eps}]{figures/ssm_blocksync.svg}
    \includegraphics[width=0.5\textwidth]{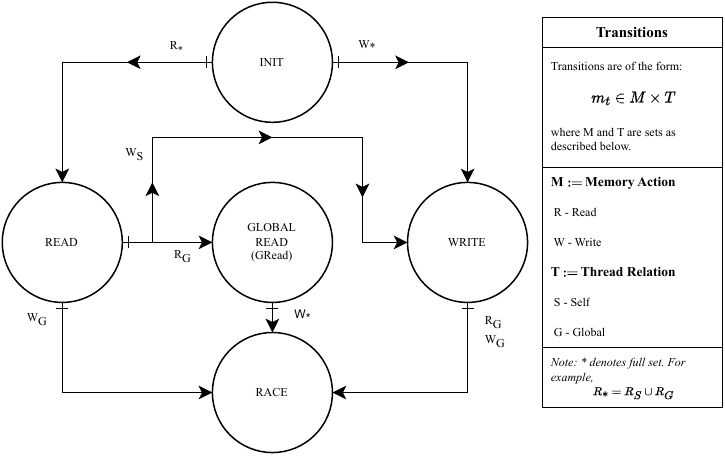}
    \caption{State machine for code without barriers; any unmentioned outgoing transition from a state means we stay in the same state---a ``self loop''
}
    \label{fig:ssm_nosync}
\end{figure}

We claim that the \fsm{} in Figure~\ref{fig:ssm_nosync}, along with a single prior accessor ID, is sufficient to track all data races that may occur in such programs along a single control-flow path.

%\footnotemark

% \footnote{This does not account for control flow dependent data races, as discussed in section~\ref{sec:background}.}

\begin{figure}[h]
    \centering
    \includegraphics[width=0.45\textwidth]{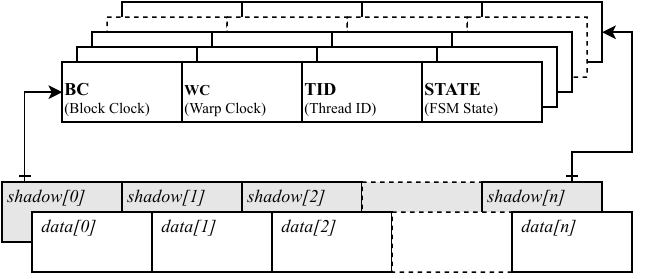}
    \caption{Shadow information used by \hirace
    }
    \label{fig:ssm_shadow}
\end{figure}

To monitor the $data$ array for races, we create a $shadow[k]$ entry 
for each array element $data[k]$,
as shown in Figure~\ref{fig:ssm_shadow}.
Each $shadow[k]$ entry maintains the following information:
the block-scalar clock ($BC$), the warp-scalar clock ($WC$), the ID of the thread performing the {\em most recent access} ($TID$),
and the state of the \fsm{} from Figure~\ref{fig:ssm_nosync} ($State$).

\begin{figure}[h]
\includegraphics[width=0.5\textwidth]{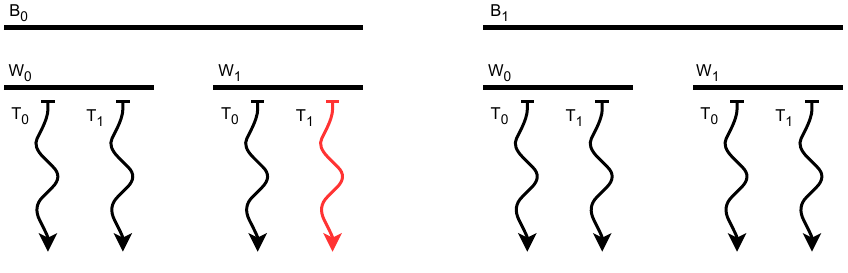}
\caption{Grid of CUDA threads assumed in Listing 1 (for which we ignore B0 and B1) and Listing 2 (for which we consider B0 and B1)}
\label{fig:simple-grid}
\end{figure}

Initially, all $State$ values are set to {\sc{init}}, indicating that no memory accesses have occurred yet.
Let us assume the thread organization 
shown in Figure~\ref{fig:simple-grid}. We label the eight threads T$_{000}$ to T$_{111}$ to indicate their block, warp, and TID. For example, the ``red'' thread is T$_{011}$.

% organized in two thread-blocks with two warps that have two threads each.
% %
% There are e

In the scenarios we describe,
all CUDA threads accessing $data[k]$ atomically
copy, due to our source instrumentation,
$shadow[k]$ into $oShadow$,
which is the tuple
$\langle 
oBC,
oWC,
oTID,
oState
\rangle$\footnote{Prefix $o$ indicates ``old'' values, i.e., the values stored by a prior accessor, whereas prefix $n$ denotes the ``new'' values determined by the current accessor.}.
Then, they determine the new tuple
$\langle 
nBC,
nWC,
nTID,
nState
\rangle$ to write back to $shadow[k]$ in a lock-free manner using an {\sc{atomicCAS}} instruction.
Only one thread will succeed; the ``losers''
re-obtain a copy of $shadow[k]$
and try again.
To illustrate this process, let us walk through a few specific scenarios.

\paragraph*{Scenario 1}
Suppose T$_{011}$ (the ``red'' thread) and T$_{100}$ concurrently execute Line 3 in Listing~\ref{lst:rw-nosync}. Suppose further that T$_{011}$ is the winner. At this point, the instrumented code transitions the $State$ from {\sc{init}} to {\sc{read}} (the transition is labeled $R_{*}$, meaning ``any read by any thread''). Since T$_{100}$'s update attempt fails, it re-reads $State$ and finds it to be {\sc{read}} with a different TID. Thus, it advances the state to {\sc{gread}}---meaning \textbf{G}lobal read---since this memory location was previously read by a different thread. If thread T$_{011}$ executes Line 4 next, the state machine follows the $W_{*}$ arc (meaning ``any write at all''), transitioning the state to {\sc{race}}.

Note that there is no obligation for the instrumented code handling $T_{100}$ to update $shadow[k]$ before T$_{011}$ executes Line 4. In other words, {\em there is no requirement that a location access and its instrumented code execute atomically}.\footnote{This is also important for performance.}

Of course, it is quite possible that only T$_{011}$ has updated $shadow[k]$ when T$_{011}$ reaches Line 4, in which case it updates the state from {\sc{read}} to {\sc{write}} following the $W_{S}$ transition, which stands for \textbf{S}ame-thread write. If the instrumented code for T$_{100}$ now updates $shadow[k]$, it enters the {\sc{race}} state since any read or write access by a different thread is a race.

\begin{figure*}[h]
    %\centering
    %\includesvg[width=\textwidth,inkscapelatex=false,extractformat={pdf,eps}]{figures/ssm_blocksync.svg}
    \includegraphics[width=\textwidth]{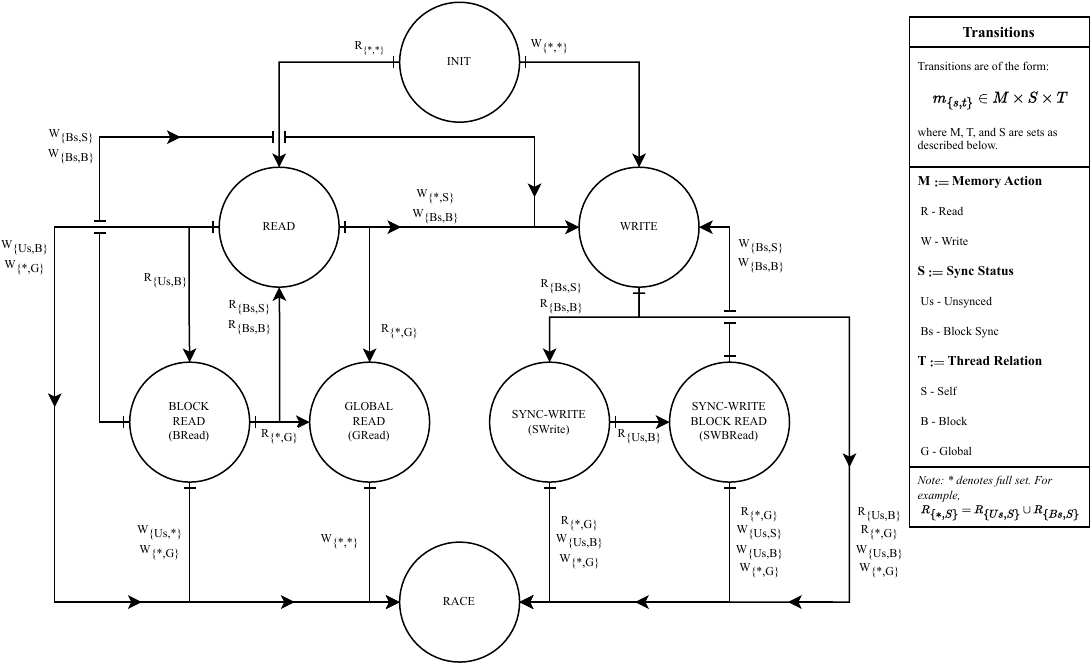}
    \caption{State-machine for code with block synchronization; any unmentioned outgoing transition from a state means we stay in the same state
}
    \label{fig:ssm_bsync}
\end{figure*}

\begin{minipage}{\linewidth}\begin{lstlisting}[caption={Read-write data race with syncthreads (block-scope barrier)},label={lst:rw-bsync}]
__global__ void race_blockSync(int* data)
{
  int tid = threadIdx.x;
  int val = data[0];
  __syncthreads();
  if (tid > 0) {
    data[tid - 1] = tid;  // race!
  }
}
\end{lstlisting}
\end{minipage}

\paragraph*{Scenario 2}

For this scenario, we consider the program in Listing~\ref{lst:rw-bsync}, which contains a block-scoped syncthreads barrier.
To explain the state-machine activities in the presence of barriers, we extend the notation on how reads and writes are subscripted.
The subscripts are of the form $\{S,T\}$, where $S$ is the synchronization status, which is either $Us$ (``unsynchronized'') or $Bs$ (``block-synchronized''), and $T$ is the thread relation, which now has a $B$ (``block'') option in addition to the former $S$ (``self'') and $G$ (``global'').
The new thread relation $B$ indicates that a barrier has ``block-synchronized'' the access (indicated by differing block scalar clock, $BC$, values).\footnote{The size of these clocks is configurable, and race detection is discontinued with a warning if a clock overflows (after reporting any previously identified races).}
The wildcard ``$*$'' means any applicable letters.
%
%We will explain the semantics of these subscripts in the scenarios below.

Consider the scenario of T$_{011}$ (the ``red'' thread of Fig.~\ref{fig:simple-grid}) executing Line 4 in Listing~\ref{lst:rw-bsync}, where it reads location $0$ of the data array. This access updates the state in $shadow[0]$ to {\sc{read}}, as indicated by the wildcard-subscripted $R_{\{*,*\}}$ transition in Fig.~\ref{fig:ssm_bsync}.
Suppose T$_{000}$, another thread in the same block, now executes Line 4. This access updates $shadow[0]$ to {\sc{BRead}}
(Block Read)---a new state---that captures the fact that two threads from the same block scope have read this location asynchronously, as indicated by the label
$R_{\{U_s,B\}}$. As long as other threads from the same block execute Line 4, $shadow[0]$ remains in the {\sc{BRead}} state.
As soon as a thread from a different block, for example T$_{110}$, executes Line 4, we promote $shadow[0]$ to {\sc{GRead}} (``global read'') as a read has been performed in the ``global scope'' ($R_{\{*,G\}}$).

Once we are in the {\sc{GRead}} state, any type of write by any thread ($W_{\{*,*\}}$) results in a {\sc{race}} as this memory location has witnessed a read from at least two blocks. This is important because syncthread barriers do not synchronize threads across blocks, so any writing thread must be executing asynchronously to at least one of the prior readers.

Instead of thread T$_{110}$ performing the aforesaid access, let $shadow[0]$ remain in {\sc{BRead}} and let the red thread T$_{011}$ execute the \verb|__syncthreads()| call on Line 5.
Because of the barrier semantics, all threads must cross the barrier before they can execute any following instruction. 
Therefore, eventually, the thread-private block-scalar clock (BC) updates to $1$ in all threads in this block.
At that point, a write by any thread in block B$_{0}$ sends the line to the {\sc{Write}} state as indicated by the transitions $W_{\{B_s,S\}}$ %in case the last-accessor itself has advanced the clock,
and $W_{\{B_s,B\}}$. 
Even so, T$_{110}$ eventually performs a read of this location, thus transition to the {\sc{Race}} state.
The remaining transitions of the \fsm{} in Fig.~\ref{fig:ssm_bsync} may be similarly explored by considering arbitrary interleaving of supported operations.

\subsection{Further Extension}

In a similar manner, the \hirace{} \fsm{} is extended beyond what is shown in Fig.~\ref{fig:ssm_bsync} to support atomics and warp-scoped synchronization primitives. This expands the full state machine to 25 states and 1200 transitions.

Atomic operations are represented as a third class of memory action (beyond Read and Write) and require several new states that describe the corresponding access histories (for example, a synchronized atomic access is distinct from a synchronized write access).

Warp-scoped barriers similarly require expanding the transition function with a new synchronization status and new thread relations. They also require additional states to capture the warp-scoped synchronization history while retaining the higher-scoped synchronization history.

This section only provides an informal view of how our \fsm{} works. To formally prove its correctness, we verified it with an exhaustive state-space search using a Murphi~\cite{Dill1996TheMV} model.

\section{Implementation}
\label{sec:impl}
% \ggcmt{JJ: Engineering details - Elaboration of SEc3 wrt impl details}

\hirace{} is implemented primarily as a template library, supported by Clang's AST tools for automated instrumentation. Its configuration is exposed through a \emph{toml} interface, allowing the user to specify a number of features, such as which kernels and variables to monitor for races, which thread groups to monitor, and bounds for shadow value metadata to reduce memory overhead.

%As discussed in Section~\ref{sec:background}, \hirace{} must monitor shared memory accesses and then determine whether any are concurrent.

%%
\subsection{Memory Access Monitoring}

%\mbcmt{``pointers'' are unclear; what about shared non-pointer variables?}
% Great point here, thanks for clarifying this for me. -john
To capture memory accesses, \hirace utilizes a templated wrapper class to monitor user data structures. For each shared memory variable, a shadow value data structure is allocated.
Both the original shared memory variable and the shadow value pointer are tracked within a thread-local instance of the \hirace{} wrapper. The wrapper then overrides relevant operators (for example, the subscript or array index operator \verb|operator[]|) to intercept memory access events and update the associated shadow value transparently.
This approach also requires overriding other relevant functions, such as atomic functions and synchronization primitives.

This methodology allows flexible management of the underlying test data structures as well as their associated shadow structures and thread metadata.
Individual threads or test data addresses may have their instrumentation enabled or disabled dynamically, and the memory footprint can be reduced by tracking only representative threads for user-defined symmetric work groups instead of tracking all threads.

\begin{algorithm}
\begin{algorithmic}[1]
\Require \\
$sAddr: \text{Shadow value address}$ \\
$access: \text{Access type (read/write/atomic)}$ \\
$tid: \text{Global thread ID}$ \\
$bc: \text{Block-scope scalar clock}$ \\
$wc: \text{Warp-scope scalar clock}$ 
\Procedure{UpdateShadow}{$sAddr, access, tid, bc, wc$}

    \Repeat

    \State $oShadow \gets \Call{atomicRead}{sAddr}$\\
    
        \State $\langle oState,$
        \State $oTid,$
        \State $oBC,$
        \State $oWC\rangle \gets \Call{UnpackShadow}{oShadow}$ \\

        \State $tRel \gets \Call{compareTids}{tid, oTid}$
        \State $sRel \gets \Call{checkSync}{bc, oBC, wc, oWC}$ \\

        \State $trans \gets \Call{getTrans}{oState, access, sRel, tRel}$
        
        \State $nState \gets \Call{StateMachineLookup}{trans}$\\
        
        \State $nShadow \gets \Call{PackShadow}{nState, tid, bc, wc}$    
    \Until{$\Call{atomicCAS}{sAddr,oShadow,nShadow}$}

\EndProcedure
\end{algorithmic}
\caption{Lock-Free Shadow Value Update}
\label{algo:updateShadow}
\end{algorithm}

\subsection{Race Detection}

The core of \hirace's detection logic is embedded in the shadow state update algorithm (Algorithm~\ref{algo:updateShadow}) built around the state machine described in Section~\ref{sec:design}.

As previously described, our \fsm{} contains 25 states and 1200 transitions. It requires only 5 bits of memory to record the \fsm{} state per monitored memory address, which is combined with other metadata (as described in Fig.~\ref{fig:ssm_shadow}) resulting in 8 bytes of memory per address by default (though this is configurable and can be reduced). In contrast, the state of the art uses 16 bytes per monitored address to track 2 prior accessors with no ability to reason about any earlier accesses in general.

% \hirace supports:
% \begin{enumerate}
%     \item Three levels of thread hierarchy (i.e., grid, thread-block, and warp).
%     \item All CUDA synchronization primitives
%     \item Atomic memory accesses.
% \end{enumerate}
%
% We have verified our state machine with an exhaustive state-space search in Murphi\cite{Dill1996TheMV}, which proved that arbitrary access patterns (within the supported set of operations) accurately map to race or non-race states.

When a thread accesses a monitored memory address through the \hirace{} wrapper, it calls Algorithm~\ref{algo:updateShadow}. This causes the thread to (atomically) check the metadata for the memory location it is accessing, compare it against its own current metadata, and use the result to update the state to a new shadow value (again atomically).

In more detail, the first step is entering the lock-free loop in Line 7 of Algorithm~\ref{algo:updateShadow}.
Any accessing threads may concurrently compare the old shadow value ($oShadow$) to their own metadata, but a thread may only store its new shadow value ($nShadow$) if it verifies, via the \verb|atomicCAS|, that $oShadow$ has not changed.
If any other thread updates the shadow value in the meantime, the current thread detects the change, starts over, and re-calculates $nShadow$ relative to the newly read $oShadow$.
Thus, all shadow-value accesses in shared memory are atomic and guaranteed to be single transitions represented by our \fsm.

Since most of the race-detection logic is contained within the \fsm, the amount of computation in the \verb|atomicCAS| loop is minimal.
Each shadow value is a bit array as outlined in Fig.~\ref{fig:ssm_shadow}.
$UnpackShadow$ splits the shadow value into individual ``old'' metadata values from the prior access.
$CompareTids$ and $CheckSync$ compare both old and new metadata values to determine the Thread Relation and Sync Status labels as described in Fig.~\ref{fig:ssm_bsync}.

The \hirace{} \fsm{} is encoded as a flat array indexed by a concatenation of $oState$ and the transition labels $m_{\{s,t\}}$ shown in Fig.~\ref{fig:ssm_bsync}, which map to a new state.
$GetTrans$ creates an integer index into the state-machine array using bit operations to combine $oState$ with the current Memory Action ($access$), Thread Relation, and Sync Status.
Afterwards, $StateMachineLookup$ is just an array access to index $trans$.

Finally, the newly calculated state is concatenated with the current metadata, and the thread attempts to store the result.

\subsection{Source Instrumentation}

To facilitate instrumentation with our wrappers, we provide a standalone Clang tool for AST parsing and rewriting. 
The Clang LibTooling C++ interface facilitates parsing source code into AST as well as walking and rewriting the AST and/or source code through a robust API.

As \hirace{} is a header library designed to wrap around existing data structures and override relevant CUDA language primitives, it can be utilized through manual instrumentation fairly easily.
However, the \hirace{} Clang tool automates this instrumentation. The Clang AST allows us to mechanically identify and wrap monitored data pointers and to allocate and deallocate necessary metadata.
Users may simply execute this tool on their source code to generate instrumented source,  which may be compiled and executed immediately or analyzed by hand or static tools if necessary.

% \begin{itemize}
%     \item Expt methodology, machines, instr methods,

%     \item Eval case studies:

%     \begin{itemize}
%         \item Irreg codes studied: gen nature

%         \item reg codes 
        
%     \end{itemize}

% \item Tools we compare against: iGuard, CUDA memcheck

%  \begin{itemize}
%      \item how we bettered over iGuard in terms of litmus tests. Informs of the possibility of adaptations we will make.

% \item ThreadFence wrong in iGuard and how we handle it

% \item iGuard overheads due to Jitting
     
%  \end{itemize}
 
% \item Kinds of measurements we will do, parameters of the examples that are relevant to race checking (threads, blocks,...) that we will vary

% \end{itemize}

% {\em .75 pages into IMPL}

% To avoid data movement and communication with the host machine, race detection is takes place entirely on the GPU. This allows us to utilize resources and patterns employed in the test code for a natural parallel implementation with each thread logging its own access histories. To achieve this, \hirace is implemented as a Clang source rewriting front end along with a header library containing the safety state machine description and necessary metadata analysis functions.

\section{Evaluation}
\label{sec:eval}

%\ggcmt{Huge deal about the exp methodology that comes with calibrated race gen methods}

%\ggcmt{TODO: put in simpler races that show iGuard omission and MemCheck missing shared-mem races.}

\subsection{Experimentation Methodology}

Our experiments were executed on a system with the following configuration:

\begin{itemize}
    \item CPU: AMD Ryzen 5 3600
    \item GPU: NVIDIA Geforce RTX 2070 Super
    \item OS: Ubuntu 22.04
    \item CUDA version: 11.7
\end{itemize}

\subsection{Section: Indigo Benchmarks}
%\ggcmt{mb describes the "calibrated benchmark generator" here}
We use programs and inputs from the recently released Indigo benchmark suite~\cite{mb-ispass22} to drive our experiments. Unlike conventional suites, Indigo contains scripts and configuration files that allow the user to generate the desired codes and inputs. 

There are 14 graph generators to choose from for creating parameterizable inputs of arbitrary size. There are also 21 CUDA kernel generator patterns, which generate up to 580 distinct CUDA programs operating on a given input data type.

More importantly for us, the code generator is not only able to synthesize hundreds of different versions of common graph-processing code patterns but also to systematically insert various bugs. We make extensive use of this latter capability as it enables us to control precisely which bugs, if any, are present. This provides a reliable ground truth valuation, which is typically not available with other suites or codes. 
Each benchmark kernel comes paired with a sequential implementation of the same algorithm as a baseline verification of the benchmark's correctness (or lack thereof).

%\mbcmt{maybe mention how many CUDA codes in total, how many buggy codes, and how many inputs were generated and used}

\subsection{Results}

%o\ggcmt{JJ will describe Table 2 roughly here. Also Fig4 might be replaced by a table/?}

\begin{table*}[t]
% \begin{large}
\begin{tabular} {lccccccccc} % {lrrrrrrrr}
\hline 
 \shortstack{Input Graph} &
 \shortstack{Number of\\ tests} &   
 \shortstack{\hirace \\ Races \\ Found} &
 \shortstack{\hirace \\ Races \\ Missed} &
 \shortstack{\iguard \\ Races \\ Found} &  
 \shortstack{\iguard \\ Races \\ Missed} &  
 \shortstack{Compute \\ Sanitizer \\ Racecheck \\ Races Found} &  
 \shortstack{Compute \\ Sanitizer \\ Racecheck \\ Races Missed} &    
 \shortstack{Sequential\\ Comparison \\ Races Found} &  
 \shortstack{Sequential\\ Comparison \\ Races Missed} \\
\hline
 DAG\_100n\_200e         &    580 &     346 & 0   &  298 & 48  &    92 & 254  & 301 & 45   \\
 DAG\_200n\_400e         &    580 &     346 & 0   &  298 & 48  &    92 & 254  & 301 & 45   \\
 DAG\_5n\_5e             &    580 &     182 & 164 &  118 & 228 &    92 & 254  & 44  & 302  \\
 counterDAG\_200n\_1000e &    580 &     346 & 0   &  298 & 48  &    56 & 290  & 282 & 64   \\
 counterDAG\_5n\_5e      &    580 &     295 & 51  &  148 & 198 &    92 & 254  & 194 & 152  \\
 power\_law\_200n\_1000e &    580 &     346 & 0   &  297 & 49  &    128 & 218 & 306 & 40   \\
\hline
\end{tabular}
% \end{large}
\caption{\normalfont Number of Indigo benchmark codes tested by each tool on the listed input. \\
Here, the column ``Sequential Comparison''
indicates benchmarks for which sequential
execution and parallel execution did not produce matching results---indicating that the parallel execution exhibited a data race.
These results are discussed further in Section~\ref{sec:discussion-of-results}. }
\label{tab:errors-found}
\end{table*}

For our tests we, used Indigo to generate 580 CUDA kernels, of which 346 contained data races. We also generated a number of small input graphs for testing the accuracy of \hirace{}, and used 6 graphs ranging between 5 and 200 nodes to evaluate both \hirace{} and \iguard. The results are displayed in Table~\ref{tab:errors-found}.

We also compared our results to \cudamemcheck's {\sc racecheck} tool, but since it only checks for CUDA block-shared-memory races, there were relatively few examples to compare against, and both \iguard and \hirace{} found all block-shared-memory races identified by \cudamemcheck.

We found that, on the 3480 Indigo benchmark comparisons between \iguard and \hirace{}, neither tool provided any false positives. While \iguard performed quite well in identifying data races given the constraints of classical race detection algorithms, \hirace{} was able to find 100\% of the injected data races in the Indigo benchmarks (when provided with inputs that exhibited the injected data race).

\begin{figure}[]
    %\centering
    \includegraphics[width=\linewidth]{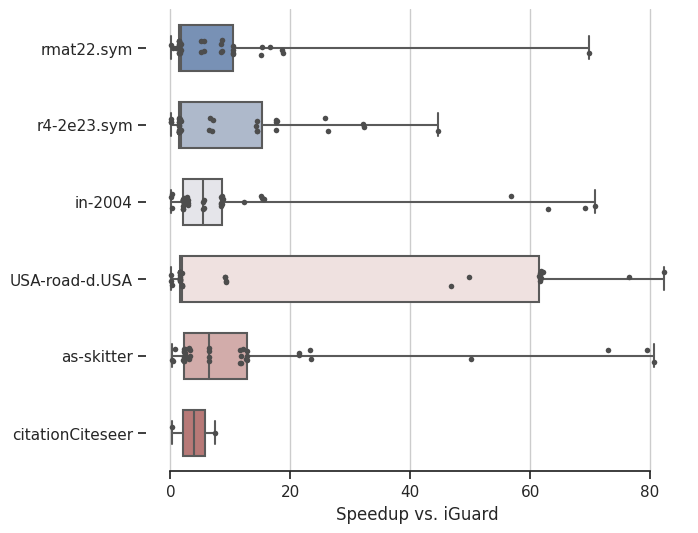}
    \caption{HiRace speedup vs. iGuard. Each point represents execution of one Indigo benchmark kernel on the associated input graph.}
    \label{fig:speedup_iguard}
\end{figure}

Further, we tested the same codes on eleven large scale real-world graph inputs (as described in Table~\ref{tab:large-graphs}) to test performance, as shown in Figure~\ref{fig:speedup_iguard}.
\iguard struggled with large grid dimensions, and the timing results were very sporadic (occasionally requiring multiple hours, yielding over 1000x slowdown over baseline code execution).
We were able to compare 441 executions across these large graphs to \iguard, which demonstrated an overall average overhead of more than 30x slowdown, against \hirace{}'s average of less than 3x slowdown across the same code/input combinations.

\begin{table}[]
\begin{tabular}{llrr}
\hline
Graph             & Type                  & Vertices & Edges    \\ \hline
rmat16.sym       & RMAT                  & 65,536    & 967,866   \\  
internet         & Internet topology     & 124,651   & 387,240   \\ 
USA-road-d.NY    & road map              & 264,346   & 730,100   \\ 
citationCiteseer & publication citations & 268,495   & 2,313,294  \\ 
amazon0601       & product co-purchases  & 403,394   & 4,886,816  \\ 
2d-2e20.sym      & grid                  & 1,048,576  & 4,190,208  \\ 
in-2004          & web links             & 1,382,908  & 27,182,946 \\ 
as-skitter       & Internet topology     & 1,696,415  & 22,190,596 \\ 
rmat22.sym       & RMAT                  & 4194304  & 65,660,814 \\ 
r4-2e23.sym      & random                & 8,388,608  & 67,108,846 \\ 
USA-road-d.USA   & road map              & 23,947,347 & 57,708,624 \\ \hline
\end{tabular}
\caption{Real-world graph inputs}
\label{tab:large-graphs}
\end{table}

% \begin{table*}[]
% \begin{tabular}{|l|l|l|l|l|}
% \hline
% name             & type                  & vertices & edges    & size {[}MB{]} \\ \hline
% rmat16.sym       & RMAT                  & 65536    & 967866   & 8             \\ \hline
% internet         & Internet topology     & 124651   & 387240   & 3.6           \\ \hline
% USA-road-d.NY    & road map              & 264346   & 730100   & 6.9           \\ \hline
% citationCiteseer & publication citations & 268495   & 2313294  & 10.3          \\ \hline
% amazon0601       & product co-purchases  & 403394   & 4886816  & 21.2          \\ \hline
% 2d-2e20.sym      & grid                  & 1048576  & 4190208  & 37.7          \\ \hline
% in-2004          & web links             & 1382908  & 27182946 & 114.3         \\ \hline
% as-skitter       & Internet topology     & 1696415  & 22190596 & 95.5          \\ \hline
% rmat22.sym       & RMAT                  & 4194304  & 65660814 & 542.1         \\ \hline
% r4-2e23.sym      & random                & 8388608  & 67108846 & 570.4         \\ \hline
% USA-road-d.USA   & road map              & 23947347 & 57708624 & 557.5         \\ \hline
% \end{tabular}
% \caption{Real-world graph inputs}
% \label{tab:large-graphs}
% \end{table*}

With \hirace{}'s low overhead, we were able to more thoroughly test its performance against baseline codes. Across 5,105 executions (464 Indigo codes across the 11 large graphs in Table~\ref{tab:large-graphs}), \hirace{} demonstrated 7.5x average slowdown, and 1.08x median slowdown (while still identifying 100\% of the injected data races, with no false positives).

\hirace{} also has a few outliers in runtime, primarily due to its race reporting method, which requires significantly more I/O on some code patterns to report races on unique memory addresses. We intend to address this (and provide configuration options for reporting) in the future.

\begin{figure}[]
    %\centering
    \includegraphics[width=\linewidth]{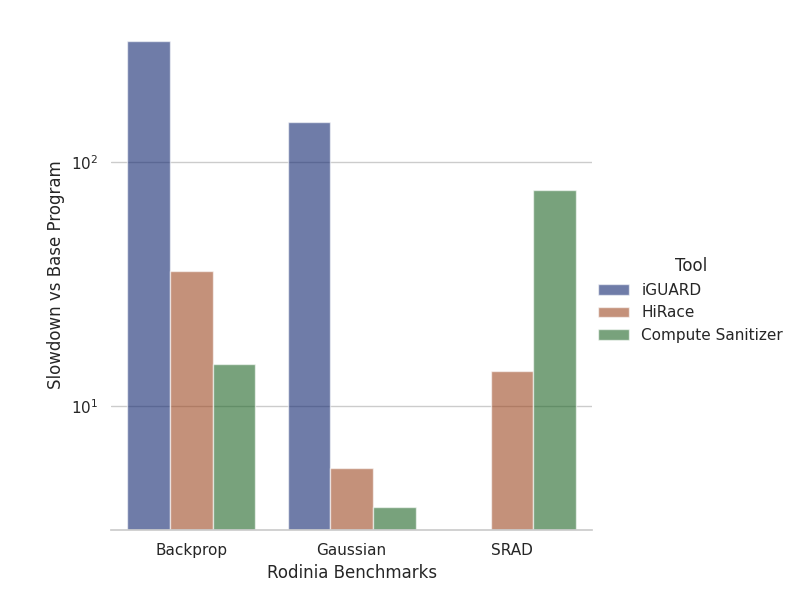}
    \caption{Race-Detection Slowdown of Rodinia Benchmarks.}
    \label{fig:rodinia_slowdown}
\end{figure}

For additional perspective, we compared the same three tools on some\footnote{The Rodinia codes are old, and most of them contain features that are no longer supported by recent versions of the nvcc CUDA compiler.} benchmarks from the Rodinia~\cite{che2009rodinia} suite. Rodinia contains a number of CUDA codes representing common GPU code patterns.
The tested programs and runtime performance are described in Figure~\ref{fig:rodinia_slowdown}.

SRAD~\cite{10.1109/TIP.2002.804276} (Speckle Reducing Anisotropic Diffusion) is a diffusion algorithm for ultrasonic and radar imaging applications.
Although it utilizes relatively little block-scoped shared memory, \cudamemcheck is several times slower than \hirace, exhibiting more than 76x slowdown of the base program. \hirace{} by comparison has less than a 14x slowdown.
\iguard, unfortunately, is unable to run this code.

Backprop is a machine learning back-propagation algorithm used to train neural-network layers. This algorithm consists of forward and backward phases updating a network of user-defined depth.
\cudamemcheck{} performs much better on this program with less than 15x slowdown. \hirace{} incurs 36x slowdown, whereas \iguard{} has over 300x slowdown. We believe this is due to \iguard's use of CUDA's Unified Virtual Memory (UVM).

The Gaussian program is a simple Gaussian Elimination algorithm for 2D matrices. 
\cudamemcheck{} incurs less than 4x slowdown as the program does not utilize block-shared-memory, so there is nothing for it to monitor.
\hirace{} is able to check global memory usage with less than 6x slowdown.
\iguard{} also struggles to scale with higher matrix dimensions on this kernel, showing over 145x slowdown.

Lastly, using our default shadow value footprint of 8 bytes per tracked memory address, we performed all of our tests with less than half of the memory overhead of \iguard. In a real application, we would be able to reduce this footprint further by configuring \hirace{}'s settings to more efficiently track the parallel patterns of a given test application.

\subsection{Discussions}
\label{sec:discussion-of-results}

One of the largest weaknesses of dynamic race detection tools is the need for arbitrary length access histories.
Maintaining a full access history is infeasible, so these tools must balance the history length against memory overheads.

% \noindent To verify the accuracy of \hirace{}, we investigated a number of patterns that are problematic for modern tools. We outline a few here by way of simple litmus tests, which we found to evade detection by \iguard, though there are still other data races missed by \iguard that we have not yet classified.

\begin{figure} \centering
\begin{minipage}{.9\linewidth}
\begin{lstlisting}[caption={Finite-history litmus test},label={lst:litmus-finiteHistory}]
__global__
void multiRead_Race(int* data, int len)
{
  int tid = threadIdx.x;
  int local_sum = 0;
  
  if (tid < len - 2) {
    local_sum = data[tid]; // evicted
    local_sum += data[tid + 1];
    
    data[tid+1]=local_sum; // race!
  }
}
\end{lstlisting}
\end{minipage}
\end{figure}
%input{figures/litmus_finiteHistory}

Listing~\ref{lst:litmus-finiteHistory} demonstrates a simple pattern in which 2 distinct threads read the same location in succession.
Though some CPU race detection tools can afford longer histories, which may capture both reads, \iguard{} in particular relies on a single prior reader and a single writer (due to the higher cost of GPU thread metadata).
Because of this, \iguard{} fails to identify any data race in Listing~\ref{lst:litmus-finiteHistory}'s kernel.

% The first, and we suspect most impactful, pattern that causes missed races is one in which multiple threads read the same memory prior to a write access by the last reader (see Listing~\ref{lst:litmus-finiteHistory}.) 
%
For example, a read from $data[1]$ by thread 1 on Line 8 is lost if thread 0 later reads the same location on Line 9.
Thus, when thread 0 writes to $data[1]$ on Line 10, although the access is concurrent to the read by thread 1 on Line 5, this history has been lost (in this particular thread schedule) and the race goes undetected. Note that this example can easily be extended to any finite history depth, thus avoiding detection on any prior race checker that is based on a finite history.

With \hirace's \fsm{} design, any number of reads and any scheduling may be represented entirely within the {\sc{Read}}, {\sc{BRead}}, {\sc{GRead}}, and {\sc{Write}} states and must eventually reach the {\sc{Race}} state.

\noindent In regards to accuracy, there are many takeaways
from Table~\ref{tab:errors-found}:
\begin{compactitem}
\item We tested these tools 
over $580\times 6 = 3480$ tests with 2076 injected races. \hirace caught all but 215 of these races---a ``{\em completeness rate}'' of 93\%.

\item All races missed by \hirace happened in executions with the smallest test graph used (5 nodes and 5 edges), where it is likely that the control-flow does not reach the data race. As discussed in Section~\ref{sec:intro}, Indigo's irregular code patterns are particularly well suited for testing this weakness of dynamic analysis tools.

\item The ``Sequential Comparison'' results indicate the success rate of comparing GPU kernel results against a sequential implementation of the same algorithm.
The fact that this naive testing method yields better results than state-of-the-art detection tools is distressing but further supports the value of \hirace's methodology.

\item The reasons why other tools miss races in specific cases are unknown to us. Part of the reason is likely due to the eviction of access records as discussed above. 

\item In general, a compiler is {\em not required to compile a high-level program containing data races in any predictable manner}~\cite{jack-about-boehm-adve} (also known as ``catch-fire semantics'').
This raises doubts in our minds on whether tools that instrument compiled code are operating on valid programs when asked to check for races at the assembly level. % (a fairly circular situation).

\end{compactitem}

\section{Concluding Remarks}
\label{sec:concl-remarks}
Whereas GPUs are powering much of the high-end HPC and ML revolution, there is a crucial omission in the arsenal of tools
necessary to keep GPU programs trustworthy---namely {\em data-race checkers}.
Given the growing uses of GPUs in safety-critical applications (e.g., automobile navigation systems) and even to implement various encryption schemes (fully homomorphic encryption or zero-knowledge proofs), and given the absence of a usable race checker for NVIDIA's GPUs, our work has the potential to plug a major hole in GPU program correctness checking.

We contribute a GPU data-race checker called \hirace that is sound (no unwarranted races reported) and (for all practical purposes) complete.
All dynamic race checkers suffer from incompleteness due to schedule generation, that is, if the program contains data-dependent control flow, 
one must drive the code with sufficiently different test inputs to cover these paths.
For instance,
considering that \hirace's 
tests include   graph benchmarks,
one must, in principle, feed all graphs of all topologies to ensure 
completeness.
There have been efforts to formally derive {\em cut-off bounds}
to bound these tests~\cite{cutoff-bounds}. Unfortunately, these works are unable to consider anything close to the kinds of problems we examine here.

Another type of completeness is based on {\em predictive race checking}~\cite{bond-predictive}, where one checks for races in one thread interleaving
and predicts whether other interleavings
contain races. 
These works largely apply to {\em mutex-based} synchronization patterns and are not applicable to GPU codes at the level discussed here (it may apply to codes that implement CUDA global barriers~\cite{DBLP:conf/oopsla/SorensenDBGR16}, which is of interest to us as future work).

In conclusion, 
we have tested \hirace on 3480 programs that are calibrated so that we can measure \hirace's bug-hunting efficiency.
We will submit \hirace for artifact evaluation, thus plugging a serious hole in data-race checking of CUDA programs.

% \vspace{1ex}
% \noindent{\bf Race-Checking as part of Compilation.}
% %
% When CUDA was introduced, there was no race checker accompanying it.
% %
% The arrival of cudaMemCheck was a huge advance, but NVIDIA stopped short of offering a race-checker for global memory races---a lack we attempt to fill.
% %
% One can also see a similar story with respect to another language---namely the
% Go-language that did not initially have a race checker, but thanks to one that exists now, many Go routines that were once deployed can now be checked for races.
% %
% In short,
% thanks to the creation of ThreadSanitizer and the Go race checker, they are now part of the compilation flags of GCC (for PThread/C) and Go codes, respectively. 
% %
% We believe that \hirace's technology offers this possibility of becoming part of CUDA compilation.

\vspace{1ex}
{\bf Future Work.}
There are a number of exciting directions for future work.
First, an efficient schedule generation method can make \hirace more complete, hence more trustworthy.
Second, if subwarp-level synchronization begins being used more, support for that in \hirace would be a good direction to pursue.
Third, and most exciting of all, \hirace's technology is almost directly usable on AMD GPUs for which there is currently no race checker.
Given the number of DOE labs that are investing heavily into AMD GPUs, we believe this would be of heightened interest to the community.

\section*{Acknowledgements}
This work has been supported in part by the National Science Foundation under Awards 1955367, 1956106,
2124100,  2217154, and
2319507.

% \section*{Acknowledgment}

% The preferred spelling of the word ``acknowledgment'' in America is without 
% an ``e'' after the ``g''. Avoid the stilted expression ``one of us (R. B. 
% G.) thanks $\ldots$''. Instead, try ``R. B. G. thanks$\ldots$''. Put sponsor 
% acknowledgments in the unnumbered footnote on the first page.

% \section*{References}

% Please number citations consecutively within brackets \cite{b1}. The 
% sentence punctuation follows the bracket \cite{b2}. Refer simply to the reference 
% number, as in \cite{b3}---do not use ``Ref. \cite{b3}'' or ``reference \cite{b3}'' except at 
% the beginning of a sentence: ``Reference \cite{b3} was the first $\ldots$''

% Number footnotes separately in superscripts. Place the actual footnote at 
% the bottom of the column in which it was cited. Do not put footnotes in the 
% abstract or reference list. Use letters for table footnotes.

% Unless there are six authors or more give all authors' names; do not use 
% ``et al.''. Papers that have not been published, even if they have been 
% submitted for publication, should be cited as ``unpublished'' \cite{b4}. Papers 
% that have been accepted for publication should be cited as ``in press'' \cite{b5}. 
% Capitalize only the first word in a paper title, except for proper nouns and 
% element symbols.

% For papers published in translation journals, please give the English 
% citation first, followed by the original foreign-language citation \cite{b6}.


\bibliographystyle{IEEEtran}
\bibliography{bibfiles/ganesh,bibfiles/ganesh-nsfmed-2019,bibfiles/burtscher,bibfiles/tyler,bibfiles/references,bibfiles/john3}

% \bibitem{b1} G. Eason, B. Noble, and I. N. Sneddon, ``On certain integrals of Lipschitz-Hankel type involving products of Bessel functions,'' Phil. Trans. Roy. Soc. London, vol. A247, pp. 529--551, April 1955.
% \bibitem{b2} J. Clerk Maxwell, A Treatise on Electricity and Magnetism, 3rd ed., vol. 2. Oxford: Clarendon, 1892, pp.68--73.
% \bibitem{b3} I. S. Jacobs and C. P. Bean, ``Fine particles, thin films and exchange anisotropy,'' in Magnetism, vol. III, G. T. Rado and H. Suhl, Eds. New York: Academic, 1963, pp. 271--350.
% \bibitem{b4} K. Elissa, ``Title of paper if known,'' unpublished.
% \bibitem{b5} R. Nicole, ``Title of paper with only first word capitalized,'' J. Name Stand. Abbrev., in press.
% \bibitem{b6} Y. Yorozu, M. Hirano, K. Oka, and Y. Tagawa, ``Electron spectroscopy studies on magneto-optical media and plastic substrate interface,'' IEEE Transl. J. Magn. Japan, vol. 2, pp. 740--741, August 1987 [Digests 9th Annual Conf. Magnetics Japan, p. 301, 1982].
% \bibitem{b7} M. Young, The Technical Writer's Handbook. Mill Valley, CA: University Science, 1989.
% \end{thebibliography}
% \vspace{12pt}
% \color{red}
% IEEE conference templates contain guidance text for composing and formatting conference papers. Please ensure that all template text is removed from your conference paper prior to submission to the conference. Failure to remove the template text from your paper may result in your paper not being published.

\end{document}